\documentclass[10pt]{article}

\usepackage{hyperref}

\usepackage{amsmath}

\usepackage{cite}

\usepackage{subfloat}

\usepackage{adjustbox}

\usepackage{caption}

\usepackage{graphicx}

\usepackage{relsize}

\usepackage{amssymb}

\usepackage{float}

\usepackage[bottom=1 in,top=1.3 in]{geometry}
\usepackage{setspace} 
\usepackage{xcolor}

\begin{document}
\title{\textbf{ Interplay of Pairing Correlation and Coulomb Correlation in Boson Exchange Superconductors}}

\author{\textbf{Koushik Mandal$^\dagger$} and \textbf{Ranjan Chaudhury$^{\ddagger}*$} \\
$^\dagger$Department of Condensed Matter Physics and Material Sciences\\
S.N. Bose National Centre for Basic Sciences \\ Block-JD, Sector-III, Saltlake, Kolkata-700106, India\\
$^\ddagger$Department of Physics\\
Ramakrishna Mission Vivekananda Educational and Research Institute \\
Belur-711202, India \\
Email- $^\dagger$koushikmandal@bose.res.in, $^{\ddagger}$ranjan@bose.res.in,\\ $^{*}$ranjan.chowdhury@rkmvu.ac.in}
\maketitle
\begin{abstract}
A theoretical methodology for exploring the conventional Bardeen-Cooper-Schrieffer (BCS) pairing instability for superconductivity from a correlated normal phase for all possible degrees of
many-body correlation, has been developed. The
Gutzwiller projection scheme with a correlation
parameter was made use of in generating the BCS pairing state. A variational scheme was
thereafter implemented, leading to a self-consistent
equation for superconducting gap function. This
equation shows explicit dependence of the gap
function on the many body correlation parameter.
This `pairing-gap' and the corresponding self-consistent gap equation in
zero correlation limit, becomes identical in nature with
those of the pure (1-well)  BCS formalism, as expected and the Coulomb correlation affects the pairing significantly with the strength of correlation. The detailed consequences are being presented here.
\end{abstract}
\maketitle
\section*{Introduction}
\vspace*{0.3cm}
Since the discovery of superconductivity, in the early 20th century, human endeavour has been steadily getting oriented towards attaining superconductivity at elevated temperatures with the ultimate dream of reaching room temperature superconductivity. In the recent past, a group of experimental physicists, had claimed very high conductivity  (nearly vanishingly small resistance) and close to Meissner value of diamagnetism at ambient temperature in a silver embedded gold plate\cite{1}. Superconductors with critical temperature $T_{C}$, above 30K,  in general belong to high temperature superconductors or high $T_{C}$ superconductors(HTSC). According to the Eliashberg theory\cite{2,3}, and its simplified version by McMillan\cite{4}, a material with very low atomic mass, generates very high frequency phonons which couples strongly to conduction electrons and thereby can produce high value of $T_{c}$. From this, analytically it is established that at high pressure metallic hydrogen should be a superconductor with a very high $T_{C}$ around 100-240K\cite{5,6}. However, it is experimentally found that not pure metallic hydrogen but hydrogen saturated compounds like sulfur hydride\cite{7} and lanthanum superhydride\cite{8} show the superconductivity at high temperatures viz. at 203K and 260K respectively at mega-bar pressure. In cuprates system, HTSC was mainly found in the  layered structure materials like $La_{2-x}Sr_{x}CuO_{4}$, $YBa_{2}Cu_{3}O_{7-\delta}$ \cite{9,10}. But, in those materials, the mechanism behind observed anisotropic pairing is not fully understood yet.\\
The conventional isotropic (s-wave) pairing interaction for Cooper pair formation is mediated by the electron-phonon coupling and is greater in magnitude than that of the `suppressed' Coulomb interaction even by an arbitrarily small amount for certain frequency range ($\omega_{C}$; of the order of Debye cut off frequency) in almost ideal Fermi sea(IFS) . This idea of phonon mediated pairing was initially used in the microscopic pairing mechanism of superconductivity put forward by Bardeen, Cooper and Schrieffer, which is known as the microscopic BCS theory \cite{11,12}. 
\\
Here, we aim to study the BCS pairing originating from a realistic normal state. In other words, we would like to understand that if Coulomb correlation among the fermions is considered, how this novel phenomena of particle-pairing, gets modified and up to what extent of this correlation the system can sustain superconducting properties. This was earlier taken care of partly in 2-well formalism within the BCS theory\cite{4}.\\
The Coulomb correlation in a system of our interest can be introduced in two different ways:(i) correlated operator algebra: correlated fermion creation, annihilation operators are defined and then dealt the system accordingly\cite{13}; and the other  is, (ii) correlated variational state: variational states are defined in which Coulomb correlation is in-built along with the pairing correlation\cite{14}. In this correlated operator approach, the commutation relations of the operators may not be exactly identical with the uncorrelated fermionic algebra but here the creation operator is indeed the complex conjugate of the annihilation operator. In literature, the hole doped  Gossameer superconductor was studied by defining this type of correlated operators. Not only that particular Gossamer system, there are other superconducting-systems, which are handled  by defining new correlated operators.  Here, we would like to go for the second approach i.e. correlated variational states and later we define two distinct way to introduce Coulomb correlation in the system.\\  
In superfluid theory of $^{3}He$, pairing is observed in Coulomb correlated fermions\cite{15}, where the trial wave functions are used to investigate the pairing with a phenomenological two body potential.  A similar
approach has been introduced in the pairing mechanism for Coulomb correlated fermions where Gutzwiller projection
operator is introduced to take care of this correlation in the system\cite{16}.
In the first  case, model states for this system are developed at first by defining the Fermi sea in which correlation is already
introduced i.e. correlated Fermi sea (CFS)\cite{17}. The CFS is defined following the Pauli exclusion principle and all states up to the Fermi momentum $\textbf{$ k_{F} $}$ are allowed to be occupied and the rest are unoccupied. Hence the pairing takes place in a narrow regime close to the Fermi surface ($\geq$ $k_{F}$).\\
The second type of variational state is defined by the action of the pairing operator on the non-interacting FS to form the pairing state first and then the Gutzwiller operator acted on it as to block the double occupancy of a site . In other words, the Gutzwiller operator is imposed on the paired fermions state to examine, how the Coulomb correlation  competes with the pairing correlation. We would like to investigate  characteristic upper cut-off of the Coulomb correlation in terms of the Gutzwiller parameter, below which the `pairing-gap' can open up for a specified weak coupling regime in a boson exchange superconducting system.\\
This paper is configured as follows: In Sec.II we introduce  our model system along with the possible variational states. The variational correlated Bardeen-Cooper-Schrieffer(CBCS) states can be described in two different ways and we first briefly demonstrate that one of those specified states are more physical than the other. In Sec.III and Sec. IV we elaborate on mathematical calculations and discuss our results. In Appendix-A and Appendix-B, the normalization of the CBCS state and the total energy expectation expression are derived respectively.
\section*{Model system}
 The Hamiltonian for the conventional superconducting system is described by the zero temperature reduced BCS Hamiltonian\cite{11}
 \begin{equation}
 H = \sum_{k}2\epsilon_{k} b_{k}^{+}b_{k} + \sum_{k,l} V_{k,l} b_{k}^{+}b_{l}
 \end{equation}
 where, $b_{k}^{+}(b_{k})$ is the electron pair creation (annihilation) operator in the momentum states $k\uparrow$ and $-k\downarrow$;\\
 \begin{eqnarray}
  \begin{split}
 b_{k}^{+} = c_{k\uparrow}^{+}c_{-k\downarrow}^{+}; 
 b_{k} = c_{-k\downarrow}c_{k\uparrow}
 \end{split}
 \end{eqnarray}
 $\epsilon_{k}$ is the single particle energy in the momentum state k, as measured from the Fermi energy; whereas, $V_{k,l}$ is a interaction matrix element connecting  two paired momentum states k and l.\\
 We start with this Hamiltonian and then  introduce electronic Coulomb correlation in terms of the correlated variational states. This correlated variational states can be formed in two different ways, as discussed below. \\
\textbf{Case-I(a)}. In this case the correlated state is defined as the correlated fermion pairing state that is formed using the Gutzwiller projected out states with doubly occupied sites from the non-interacting filled Fermi sea(FS) ground state and it is given as  
 \begin{equation}
 \mid\Psi\rangle_{CBCS} = P_{BCS}\otimes P_{G}\mid FS \rangle
 \end{equation}
 where $P_{BCS}$ is the BCS pairing operator as defined for the construction of the  microscopic BCS pairing state and $P_G$ is the Gutzwiller partial projection operator. For the time being, we are not defining these two operators in great details, however in the subsequent sections we will discuss in details about those two operators.\\
\textbf{(b)} The effect of Coulomb correlation on pairing phenomena has also been studied for a Cooper's  one pair problem, using a variational wave function in the first quantized form i.e. Jastrow geminal augmented wave function\cite{18,19}. In this formulation, the relative distance($\rho$) between two electrons carries the effect of Coulomb correlation and the effect will be maximum when the distance is zero i.e. the electrons are on the same site. Accordingly, there is a kink in the two-electron wave function for $\rho  = 0$ and the effect of correlation weakens off gradually with the increase in separation. However, the so called bound state energy for the electron pair is affected significantly due to the Jastrow factor. Here, because of this Jastrow correlator, a repulsive matrix element will appear in the calculation of two-body interaction which  plays an important role in the calculation of the bound state energy\cite{20}.  Infact a 2-well scheme involving an attractive well arising from a boson mediated pairing interaction and a repulsive well due to Coulomb interaction was implemented for Cooper's one pair problem as well\cite{21}.
\\
Thereafter, we proposed the variational state (see Eqn(3)) in case-1(a), in second quantized form, to study the effect of correlation in 2D interacting systems. Here in this approach, we are first applying the Gutzwiller projection operator on the FS state to define the correlated FS(CFS) state. Then, we apply BCS pairing operator on the CFS states to form the correlated BCS(CBCS) states. The Gutzwiller partial projection operator is used to exclude the doubly occupied states for the HTSC antiferromagnetic cuprate superconducting system like, $La_{2-x}Sr_{x}CuO_{4}$, $YBa_{2}Cu_{3}O_{7}$\cite{22,23}.\\
\textbf{Case-II} The second approach for the correlated state is defined as follows:\\ (i) first we define a BCS pairing state i.e. BCS pairing operator($P_{BCS}$) acting on the filled FS state and then (ii) we operate  Gutzwiller projection operator($P_{G}$) on the pairing states\\
\begin{equation}
\mid\Psi\rangle_{C} = P_{G}\otimes P_{BCS}\mid FS \rangle
\end{equation}
Here, the projection operator is blocking the electrons to sit on the same site and this blocking is controlled by a variational parameter `$\alpha$' ($\alpha$ varies from 0 to 1,  where, $\alpha$ = 0 denotes the usual BCS like pairing ground state, whereas, $\alpha$ = 1 represents the complete projecting out of all the doubly occupied sites). \\
So, in the paired ground state with general $\alpha$, one has a fermionc pairing from a metallic correlated state.\\
In the first approach (case-I (a)), our derived results do not exactly match with that of the uncorrelated BCS formalism even when we consider $\alpha$ = 0. Again, in the strongly correlated regime i.e. $\alpha\geq 0.6$, the pairing gap increases slowly with the tuning parameter $\alpha$, which may be a signature of the existence of `frozen-Cooper pair' i.e. paired fermions which are almost immobile.  The incomplete variational calculation  was undertaken, which might have let to this kind of results.\\
The ratio of zero temperature pairing gap to the product of Boltzmann constant and critical temperature denoted by $\dfrac{\Delta}{k_{B}T_{c}} $, obtained was slightly smaller than that of the original BCS formulation. This may be because of the further limitation in the calculation, viz.  states with higher order corrections in $\alpha$, were assumed to be insignificant or might be the restriction on the   the filling factors on the momentum space. In the strong correlation regime ($\alpha\approx 1$), $\dfrac{\Delta}{k_{B}T_{c}} $ goes to zero or approaches to zero. Therefore the, s-wave pairing is no more feasible in this strongly correlated region, although there may be a possibility of the existence of the anisotropic (d-wave) pairing. \\
Here, in this paper, we will not elaborate on the first approach(case-I), but rather we concentrate on the second formalism and as a future plan we are willing to treat the first one as a separate problem.  In the next section we give a detailed presentation on the second approach(case-II) accombined with all the calculations and aspects. 
\section*{Mathematical formulation and construction of correlated BCS states}
Here, we start with the well known microscopic BCS pairing states and later we introduce the electronic correlation  to describe the correlated BCS states. The microscopic BCS pairing state which was modelled to capture the physics of the pairing between a pair of fermions in the presence of whole FS  is written as\\
\begin{equation}
\mid\Psi\rangle_{BCS} = \prod_{k} (u_{k} + v_{k}c^{+}_{k,\uparrow}c^{+}_{-k,\downarrow})\mid FS\rangle
\end{equation} 
where, $ u_{k}$ and $ v_{k} $ are the complex Bogoliubov amplitudes corresponding to the unoccupied paired and occupied paired states respectively and they are connected with each other by the normalization condition given by: \\
\begin{equation}
(\vert u_{k} \vert ^{2} + \vert v_{k} \vert ^{2} )= 1 
\end{equation}
Here, $\mid FS\rangle$ is the non-interacting Fermi sea ground state and we will define it later.\\
Our aim is now to introduce correlation among the pairing fermions i.e. we would like to switch on the interaction, in the system. This can be done either by (i) defining a variational correlated basis state\cite{25,26} or by (ii) introducing new correlated fermion creation, annihilation operators\cite{27,28}. The operator algebra is not exactly same as the fermion operators do. Hence, creation operator is a definite conjugate of annihilation operator but they do not follow the anti-commutation relation as followed by the fermion operators. Therefore, we are not taking this route of defining new correlated operators, rather we follow the correlated basic states technique.\\ 
Our proposed CBCS state namely that  is the Gutzwiller projected out BCS pairing states, is defined in the following way-(i) at first we define the FS ground state for the real construction of the well known microscopic  BCS ground state; (ii) then the pairing states for the correlated fermions is modelled by introducing the Gutzwiller partial projection operator in k-space which  projects out the double occupancy on a lattice site with an amplitude $\alpha$. Here, the CBCS state denoted by $\mid\Psi\rangle_{C}$ , is defined as\\
 \begin{equation}
\mid\Psi\rangle_{C} =\prod_{s,l}(1-\alpha\sum_{k^{'},m^{'}} c_{k^{'},\uparrow}^{+}c_{k^{'},\uparrow}c_{m^{'},\downarrow}^{+}c_{m^{'},\downarrow} e^{i(m^{'}-k^{'}).r_{s}})(u_{l} + v_{l} c_{l,\uparrow}^{+}c_{-l,\downarrow}^{+})\mid FS \rangle
 \end{equation}
where, $c_{k,\sigma}^{+}(c_{k,\sigma})$ is the fermion creation(annihilation) operator in momentum and spin states k and $\sigma(\uparrow,\downarrow)$ respectively. \\
 In the above equation Gutzwiller projection operator has been defined in the k-space  which takes care of the effect of exclusion of double occupancy on a single site in real space. The variational parameter $\alpha$ can be determined by minimizing the total energy expression.\\
 For the time being, the normalization factor is not being included in defining the CBCS state. We will consider this factor indeed in the subsequent sections and  we add an appendix(Appendix-A) with some relevant steps for the calculation of normalization factor.
 Generally the Gutzwiller projected out state is defined in terms of the partial projection operator in real space as given below:\\
 \begin{equation}
 \mid\Psi_{G}\rangle = \prod_{i}(1-\alpha \widehat{n}_{i\uparrow}\widehat{n}_{i\downarrow})\vert FS\rangle
 \end{equation}\\
 where, $\widehat{n}_{i,\sigma}$ is the fermionic occupation number operator at the site i with spin $\sigma$ and $\alpha$ is a variational parameter in Gutzwiller's original formulation, which decides the amplitude for double occupancy of a site ($0\leq\alpha\leq1)$; $ \vert FS\rangle $ is the Fermi sea(FS) ground state and it can be written in terms of the fermion creation operators as\\  
 \begin{equation}
 \vert FS\rangle = \prod_{k,\sigma}^{\mid k \mid \leq k_{F}}\sum_{i,j}c^{+}_{i,\sigma}c^{+}_{j,-\sigma} e^{i(r_{i}-r_{j}).k}\vert vac \rangle
 \end{equation}
  where, i and j are the site indices in real space and k represents the fermion wave vectors which has the upper bound of Fermi momentum ($ k_{F}$). $ \vert vac \rangle$ represents the vacuum state that stands for the state with zero occupation on every site in real space and  for k-space, all the momentum states being empty.\\
 For the evaluation of the ground state energy W, we combine Eqs.(1) and (7). By definition,
 \begin{equation}
 W = \dfrac{_{C}\langle \Psi\vert H\vert\Psi\rangle _{C}}{_{C}\langle \Psi\vert\Psi\rangle _{C}}
 \end{equation}
  Here, the first term in Eqn(10) corresponds to the kinetic energy operator expectation value (T) and is given by 
  \begin{equation}
  T = \dfrac{_{C}\langle \Psi\vert \sum_{k}2\epsilon_{k} b_{k}^{+}b_{k} \vert\Psi\rangle _{C}}{_{C}\langle \Psi\vert\Psi\rangle _{C}}
  \end{equation}
  This term is evaluated using the orthogonality of the independent states and it comes out as(some significant steps are put in Appendix B): 
  \begin{equation}
  T = 2\sum_{k}\dfrac{\epsilon_{k}\vert v_{k}\vert^{2}(1-\alpha)^{2}}{1+\vert v_{k}\vert^{2}\alpha(\alpha-2)}
  \end{equation}
 The two-body attractive interaction potential contribution (V) to the ground state energy is given by 
  \begin{equation}
  V = \dfrac{_{C}\langle \Psi\vert \sum_{k,l} V_{k,l} b_{k}^{+}b_{l} \vert\Psi\rangle _{C}}{_{C}\langle \Psi\vert\Psi\rangle _{C}}
  \end{equation}
 where, any two states specified with the index k and l are connected by a matrix element $V_{kl}$.\\
Simplifying we get,
\begin{equation}
V = \sum_{k,l}\dfrac{V_{k,l} u_{l}^{*}u_{k}v_{k}^{*}v_{l}(1-\alpha)^{2}}{[1+\vert v_{k}\vert^{2}\alpha(\alpha-2)][1+\vert v_{l}\vert^{2}\alpha(\alpha-2)]}
\end{equation}
\hspace*{10cm} [Recalling that, $0\leq \alpha\leq 1]$\\
Thus, the expression for the total ground state energy (W) becomes,
 \begin{equation}
 W = 2\sum_{k}\dfrac{\varepsilon_{k}\vert v_{k}\vert^{2}(1-\alpha)^{2}}{1+\vert v_{k}\vert^{2}\alpha(\alpha-2)} + \sum_{k,l}\dfrac{V_{k,l} u_{l}^{*}u_{k}v_{k}^{*}v_{l}(1-\alpha)^{2}}{[1+\vert v_{k}\vert^{2}\alpha(\alpha-2)][1+\vert v_{l}\vert^{2}\alpha(\alpha-2)]}
 \end{equation}
The `superconducting pairing gap' ($\Delta_{k}$) in this Coulomb correlated phase is defined as
\begin{equation}
\Delta_{k} = -\sum_{l}V_{k,l} \langle b_{k}^{+}\rangle
\end{equation}
Incorporating the expectation value of the pair operator in the CBCS state the Eqn (16) becomes
\begin{equation}
\Delta_{k} = -\sum_{l}V_{k,l} \dfrac{u_{l}v_{l}(1-\alpha)^{2}}{1+\vert v_{k}\vert^{2}\alpha(\alpha-2)}
\end{equation}
The total energy W and the pairing-gap $\Delta_{k}$ are both functions of the pairing amplitudes, $u_{k}$ and $ v_{k}$.  The function W can now be minimized with respect to all 3 variables viz. u, v and $\alpha$ to obtain the characteristics of the pairing correlation in the presence of the Coulomb correlation.\\\\
\textbf{(i)Coulomb correlation as a tuning parameter:} As a first step, for  simplicity, we consider the variation with respect to u and v only, treating $\alpha$ simply as free parameter. 
The variables  $u_{k}$ and  $v_{k}$ are taken here as a parametric transformation of the kind \\
\begin{equation}
u_{k} = sin\theta_{k} ;
v_{k} = cos\theta_{k}
\end{equation}
Making use of this, the Eqns (15) and (17) become
\begin{equation}
W =  2\sum_{k}\dfrac{\varepsilon_{k}(1-\alpha)^{2}cos^{2}\theta_{k}}{1+\alpha(\alpha-2)cos^{2}\theta_{k}} + \dfrac{1}{4}\sum_{k,l}\dfrac{V_{k,l} (1-\alpha)^{2}sin2\theta_{k}sin2\theta_{l}}{[1+\alpha(\alpha-2)cos^{2}\theta_{k}][1+\alpha(\alpha-2)cos^{2}\theta_{l}]}
\end{equation}
and
\begin{equation}
\Delta_{k} = -\dfrac{1}{2}\sum_{l}V_{k,l}\dfrac{(1-\alpha)^{2}sin2\theta_{k}}{1+\alpha(\alpha-2)cos^{2}\theta_{l}}
\end{equation}
These two fundamental equations now contain the parameters $\theta$ and $\alpha$.\\
\textbf{Minimization with respect to the Bogoliubov amplitudes:}\\
We minimize the total energy W with respect to the variable $\theta_{k}$ by taking 
\begin{equation}
\dfrac{\partial W}{\partial\theta_{k}} = 0 
\end{equation}
This minimization procedure leads to an effective gap equation as given below :
\begin{equation}
\varepsilon_{k} (1-\alpha)^{2} {\dfrac{sin2\theta_{k}}{cos^{2}\theta_{k}(\alpha^{2}-2\alpha+2)-1}} =  \Delta_{k}
\end{equation}
\hspace*{10cm}[making use of Eqn(20)]\\
 Now combining Eqns(20)and (22) we obtain,
\begin{equation}
\dfrac{\Delta_{k}}{\varepsilon_{k}} = \dfrac{(1-\alpha)^{2}sin2\theta_{k}}{(1-\alpha)^{2}cos^{2}\theta_{k}-sin^{2}\theta_{k}}
\end{equation}
Let us rewrite the tuning parameter $\alpha$ as 
\begin{equation}
\alpha = (1-g)
\end{equation} 
with the allowed regime for the new parameter g being also the same as that of $\alpha$ viz.  $ 0\leq g\leq 1$. Now under this transformation, the gap equation (Eqn(20)) becomes,
\begin{equation}
\Delta_{k} = -\dfrac{1}{2}\sum_{l}V_{k,l}\dfrac{g^{2}sin2\theta_{l}}{sin^{2}\theta_{l}+g^{2}cos^{2}\theta_{l}} 
\end{equation}
with
\begin{equation}
\dfrac{\Delta_{k}}{\varepsilon_{k}} = \dfrac{g^{2}sin2\theta_{k}}{g^{2}cos^{2}\theta_{k}-sin^{2}\theta_{k}}
\end{equation}\\
Combining  Eqns(25) and (26) we obtain the following equation
\begin{equation}
\sum_{k,l} V_{kl} \dfrac{sin2\theta_{l}}{(sin^{2}\theta_{l} + g^{2}cos^{2}\theta_{l})^{2}} [(1-g)(sin^{2}\theta_{k} sin^{2}\theta_{l} - g^{4}cos^{2}\theta_{k}cos^{2}\theta_{l})]= 0
\end{equation} 
The above equation obtained by extremization condition, gives two possible solutions for g and among which g = 1, is a natural root for this equation. The root g = 1, sets the ideal or the zero correlation regime for the system and indeed for this solution the ground state energy as well as the pairing gap will have the maximum magnitude. The other roots can be evaluated by solving the above Eqn(27) in detail manner.\\
We have eliminated $sin\theta_{k}$ and $cos\theta_{k}$  in terms of $\Delta_{k}, \varepsilon_{k} $ , g using  Eqns(26) and (27). Then making use of Eqn(25), we obtain the following self-consistent equation for $\Delta_{k}$ as   
\begin{equation}
 \Delta_{k} = -\dfrac{1}{2}\sum_{l}V_{k,l}\dfrac{g^{3}\Delta_{l}[\Delta_{l}^{2}(1+g^{2})^{2}+4g^{2}\varepsilon_{l}^{2}]^{\dfrac{1}{2}}}{\Delta_{l}^{2}g^{2}(1+g^{2})+g^{4}\varepsilon_{l}^{2}[(1+g^{2})+(1-g^{2})\sqrt{1+\dfrac{\Delta_{l}^{2}}{g^{2}\varepsilon_{l}^{2}}})]}
\end{equation}
The superconducting gap functions occurring in the above equation,corresponds to the Coulomb correlated fermion pair and these results are valid in both strong and weak correlation regimes. In the absence of induced Coulomb correlation i.e. g = 1(or $\alpha = 0$), this gap is identical to the non-interacting fermion pairing gap, as defined in the conventional BCS formalism. \\
The pairing potential $V_{k,l}$ is playing the crucial role here in pair formation and depending on the nature of the bosonic mediator,  we can set the interaction frequency regime in the 1-well model. Let it be done first for the phonon mediated pairing and we set the upper cut off frequency as $\omega_{c}$ , where $\omega_{c}$ $\simeq$ $\omega_{D}$; $\omega_{D}$ being the Debye frequency in the normal metallic phase. \\\\
 We first take the phonon mediated pairing case. This potential well is defined as following the BCS prescription:
\begin{align}
V_{k,l} & = - V  \hspace{1cm} \mbox {for}\hspace{1cm} -\hbar\omega_{c} \leq \varepsilon_{k}\leq \hbar\omega_{c}\nonumber\\
& = 0 \hspace{1.4cm}\mbox {otherwise }
\end{align}
Assuming isotropic (s-wave) pairing for gap function on the Fermi surface, we take,  $\Delta_{k} = \Delta_{l}$. Then performing the summation over all possible momentum states in the right hand side of Eqn(28), we obtain the following equation.
\begin{equation}
1 = \int_{-\hbar\omega_{c}}^{\hbar\omega_{c}} V N(\varepsilon) g *\dfrac{ [\Delta^{2}(1+g^{2})^{2}+4g^{2}\varepsilon^{2}]^{\dfrac{1}{2}}}{\Delta^{2} (1+g^{2})+g^{2}\varepsilon^{2}[(1+g^{2})+(1-g^{2})\sqrt{1 + \dfrac{\Delta^{2}}{g^{2}\varepsilon^{2}}})]} d\varepsilon
\end{equation}
where, N($\varepsilon$) is the single spin electronic density of states per unit volume, which is  nearly constant around the Fermi level and is approximated by $N_{0}$. The product of $N_{0} $ and V is defined as the dimensionless attractive coupling constant $\lambda$ as usual, i.e. $N_{0} V = \lambda$.\\
The integral on the right hand side of Eqn(30) is evaluated. In this context, we take an approximation viz. for $\dfrac{\Delta}{\varepsilon}\ll1 $. Thereafter, using this approximation and rearranging the integrand, Eqn(30) becomes,
\begin{equation}
1 = 2g\lambda \int_{0}^{\hbar\omega_{c}} \dfrac{[A^{2} +\varepsilon^{2}]^{\dfrac{1}{2}}}{B^{2}+\varepsilon^{2}} \hspace{0.4cm}d\varepsilon
\end{equation}
where,
\begin{align}
A^{2} = \dfrac{\Delta^{2}(1+g^{2})^{2}}{4g^{2}}\\
\mbox{and},  B^{2} = \dfrac{\Delta^{2}(3+g^{2})}{4g^{2}} 
\end{align}
Here , A and B are real quantities since $ g^{2}$ and $\Delta^{2} $ are both real for  g values being within the range ($ 0 \leq g\leq 1 $).\\
Now integrating the right hand side of  above Eqn(31)  numerically we obtain 
\begin{equation}
2g\lambda [\sqrt{A^{2}-B^{2}}tan^{-1}[\dfrac{\hbar\omega_{c}\sqrt{A^{2}-B^{2}}}{B\sqrt{\hbar^{2}\omega_{c}^{2} + A^{2}}}] -\dfrac{1}{2} log A^{2}+ log[\hbar\omega_{c}+\sqrt{\hbar^{2}\omega_{c}^{2}+ A^{2}}] = 1
\end{equation}
The first term within the above simplified result, can't not contribute for the parameter regime of g . This is because the argument of arctan function becomes imaginary for the allowed ranges of g between 0 to 1 which is not physically acceptable. The factor, $A^{2}-B^{2}$ is greater than zero only for g $ > 1 $, which is again unphysical. Therefore, from the Eqn(34), we get the solution for the superconducting gap ($\Delta_{P}$)  function as,
\begin{equation}
\Delta_{P} = \dfrac{4\hbar\omega_{c} g \hspace{0.1cm}exp\left( -\dfrac{1}{\lambda g}\right) }{(1+g^{2})}
\end{equation}
Thus, the superconducting gap arising from Coulomb correlated fermion pairs(with free Coulomb correlation parameter) is given by the above equation.  It must be kept in mind however, that the Coulomb correlation is treated here simply as a tune-able or free parameter by choosing  from the physically allowed range of g values.\\\\

\textbf{(ii)Inclusion of Coulomb correlation as a variational parameter:}  So far, the variational nature of  the Gutzwiller parameter $\alpha$ has not been taken into account. Now  we treat $\alpha$ as a variational parameter too. This is done by minimizing the total energy with respect to g (which is again equivalent to the minimization with respect to $\alpha$). This is expressed as,
\begin{equation}
\dfrac{\partial W}{\partial g} = 0
\end{equation}
The extremization procedure now leads to the following equation

\begin{equation}
\sum_k \varepsilon_{k} g^{2} sin2\theta_{k} + \dfrac{1}{2}\sum_{k,l} V_{k,l} \dfrac{g sin2\theta_{l}(sin^{2}\theta_{k}sin^{2}\theta_{l} - g^{4} cos^{2}\theta_{k}cos^{2}\theta_{l}}{(sin^{2}\theta_{l} + g^{2}cos^{2}\theta_{l})^{2}} = 0
\end{equation} 
After simplification and making use of Eqn(25) the above equation leads to the following equation
\begin{equation}
g^{2} = \dfrac{\Delta_{k}tan^{2}\theta_{k}}{\Delta_{k} -2\varepsilon_{k} tan\theta_{k}}
\end{equation}
The above equation(Eqn(38)) and the Eqn(26) are now two independent simultaneous equations involving the variable  $\theta_{k}$ and g, in which the interplaying character of the pairing correlation and the Coulomb correlation is manifestly present. Analytically, it is somewhat complicated to determine the solution. Therefore, we handled it numerically and after combining it with  $\Delta$ equation(Eqn(25)) to obtain the self-consistent gap equation. The self consistent `pairing-gap' equation is now solved for the attractive potential well as previously defined in Eqn(29). \\
The boson exchange pairing is taking place within the defined range of the attractive potential and the `superconducting pairing-gap'($\Delta_{C-P}$) in the presence of Coulomb correlation is given by \\
\begin{equation}
\Delta_{C-P} = 2\hbar \omega_{c} g \hspace*{0.1cm}exp\left(-\dfrac{1}{\lambda g}\right)
\end{equation} 
This time, the pairing gap is specifically denoted as $\Delta_{C-P}$ to indicate that both the correlations are considered actively in the pair formation. 
\section*{Results and Discussion}
In the last section, all the mathematical calculations are included where the bosonic mediator for the fermion pairing, is phonon. The mediating phonon energy spectrum is considered to be lower than the Fermi energy and here the Fermi energy is  assumed to be of the order of 1eV; while the pair-bound state energy is only few meV. We set the  range of potential well depending on the Debye temperature($\theta_{D}$). A table for a series of materials(some of them holds superconducting property), with their Debye temperature around 400K can be seen\cite{29,30}.\\
\textbf{Case-I: Pairing correlation in the presence of passive Coulomb correlation:}\\
At first, the well range is taken as 0.05 eV for the phonon mediated correlated pairing and the result is plotted here. The plotted result for $\Delta_{P}$ against the parameter g for different coupling constants $\lambda$[see fig:1].   
\begin{figure}[H]
\centering
\includegraphics[scale = 0.45]{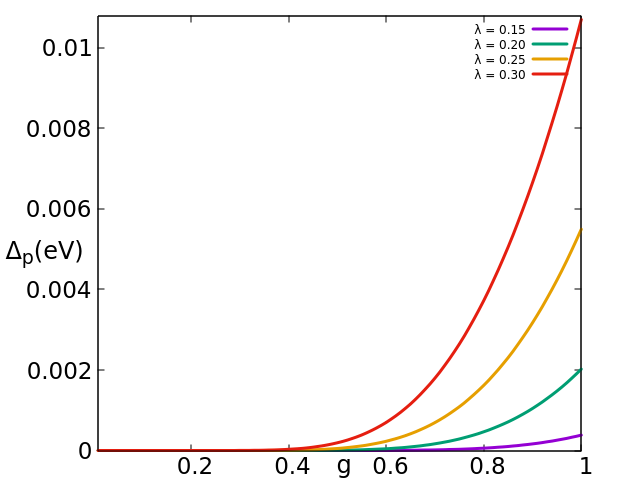}
\caption{$\Delta_{P}$ vs g for different magnitudes of coupling constant $\lambda$}.\\
\end{figure}
The result is also plotted for $\Delta_{P}$ against coupling constant $\lambda$ for different g values[see fig:2]. The pairing gap opens up, for very small but finite positive values of coupling, even for  $\alpha = 0$ (or g = 1). Thus pairing for the uncorrelated fermions in the presence of passive FS, also requires finite coupling. The colored lines are for the different coupling values $\lambda$ and the violet line is corresponding to $\lambda$ = 0.30. This curve, depicting relatively high coupling, provides the comparatively large pairing gap $\Delta_{P}$,  in the intermediate coupling regime ($\lambda \geqslant 0.30$). In the strong correlation limits the on-site Coulomb interaction dominates and confines the electrons to their real site positions. The repulsive on-site Coulomb potential dominates over the phonon mediated attractive electron-electron interaction and that significantly affects the pairing . On other hand, the weak on-site correlation allows the pair formation and the gap is effectively very small for weak coupling even with weak correlation.\\
For the weak to intermediate coupling regime viz. $ 0.3\leq \lambda\leq0.5$, the pairing gap is non-zero for relatively large correlation limit  but not for very strong correlation i.e g = 0. In that scenario, again the $\Delta_{P} $ drops down to zero for the very strong correlation and this also happens for the very strong coupling ($\lambda \rightarrow1$).\\ \\
 
\begin{figure}[H]
\centering
\includegraphics[scale=0.35]{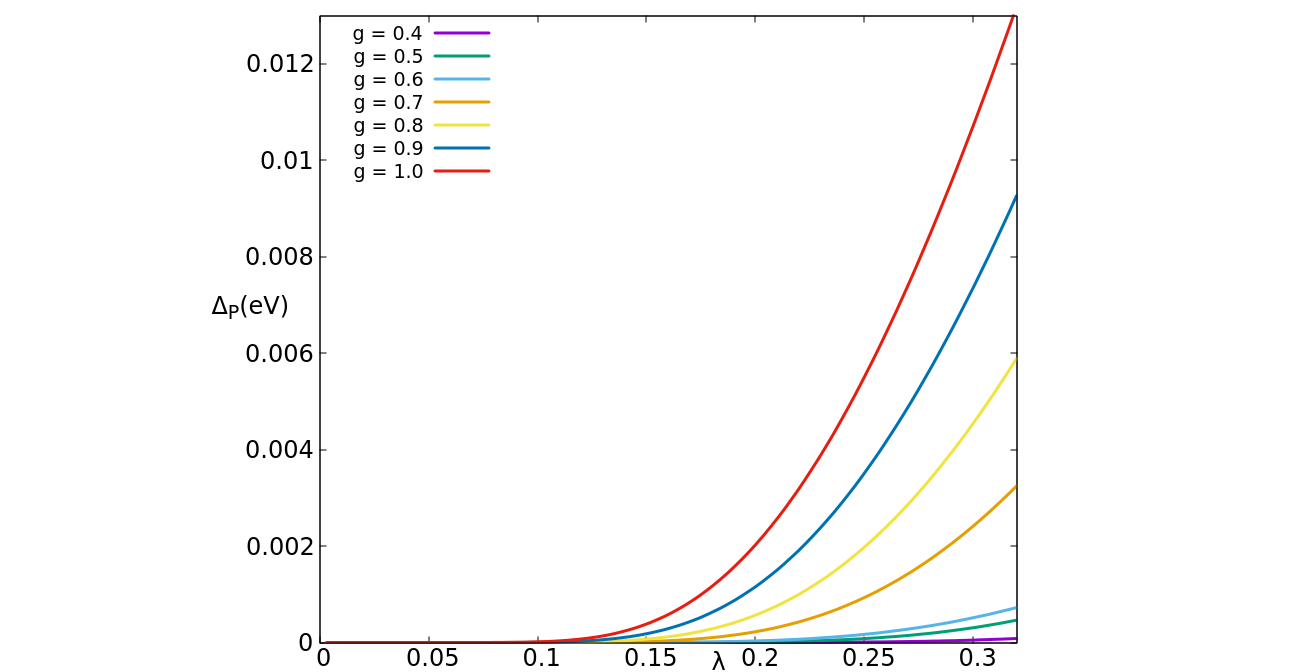}
\caption{$\Delta_{P}$  vs $\lambda $ for different g values }
\end{figure}
We have seen from fig:2 that, $\Delta_{P}$ increases with $\lambda$ for a fixed g value, as expected. In other words it implies that, with increase in the strength of coupling, pair formation is enhanced for the allowed electron states, specified by the g values. For the very low  g value (close to zero), where most of the doubly occupied states are eliminated, $\Delta _{P}$ tends to zero.  In this regime, there are no available states to form  pairs. However, pairing states will be available as  g becomes more than zero. The magnitude of  $\Delta _{P}$  will increase with `g' value and it will be maximum for g = 1, corresponding to  a specific $\lambda$ value. \\\\
\textbf{Case-II:  Pairing correlation in the presence of  active Coulomb correlation :}\\
The gap function $\Delta_{C-P}$ is first plotted  against gfor different $\lambda$ values[see fig:3] as well as  it is also plotted against  $\lambda$  for different g values[see fig:4]. The calculational results are presented here only in the weak to intermediate coupling regime.
\begin{figure}[H]
\centering
\includegraphics[scale = 0.38]{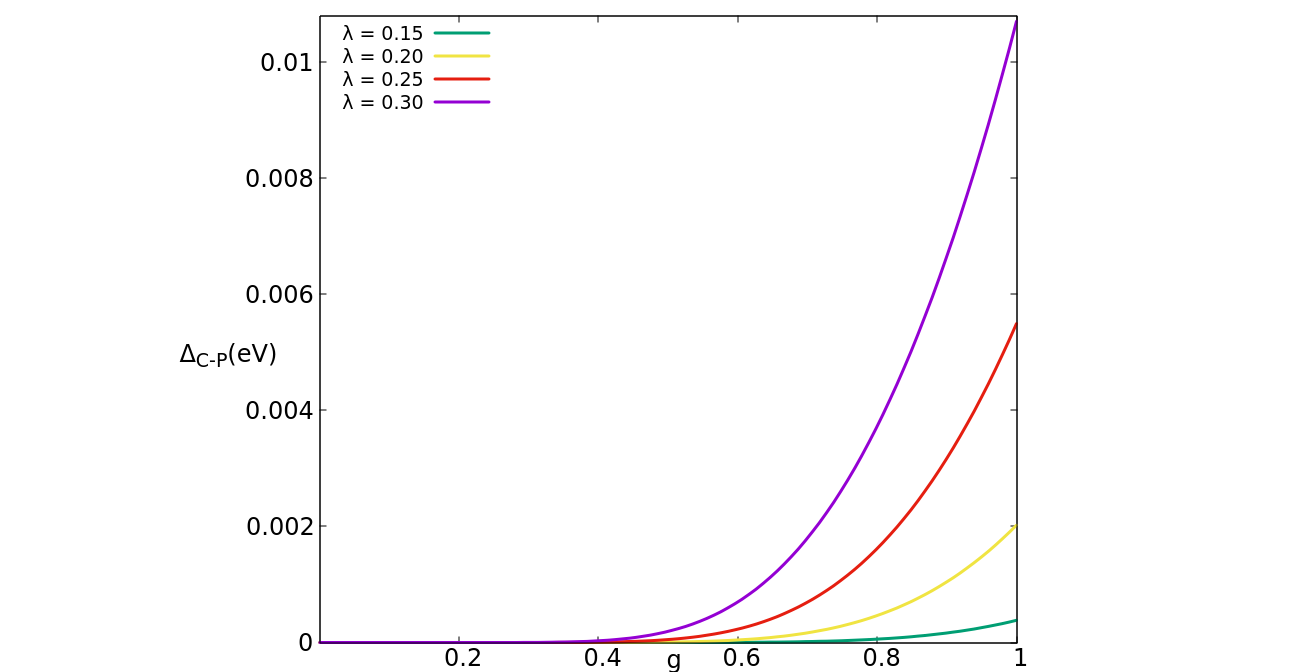}
\caption{$\Delta_{C-P}$ vs g for different magnitudes of coupling constant $\lambda$}.\\
\end{figure}
Both  $\Delta_{P}$ and $\Delta_{C-P}$ are plotted against g [see fig:1 and fig:3], and these two follow the similar exponential nature; but the variation of $\Delta_{C-P}$ with g is a little sharper for a specified $\lambda$ value. The pairing correlation is dominating over the on-site repulsion. In other word, it can be said that the electrons are not forced to stiff to their positions; rather they are relatively free to form  pair.\\
\begin{figure}[H]
\centering
\includegraphics[scale = 0.38]{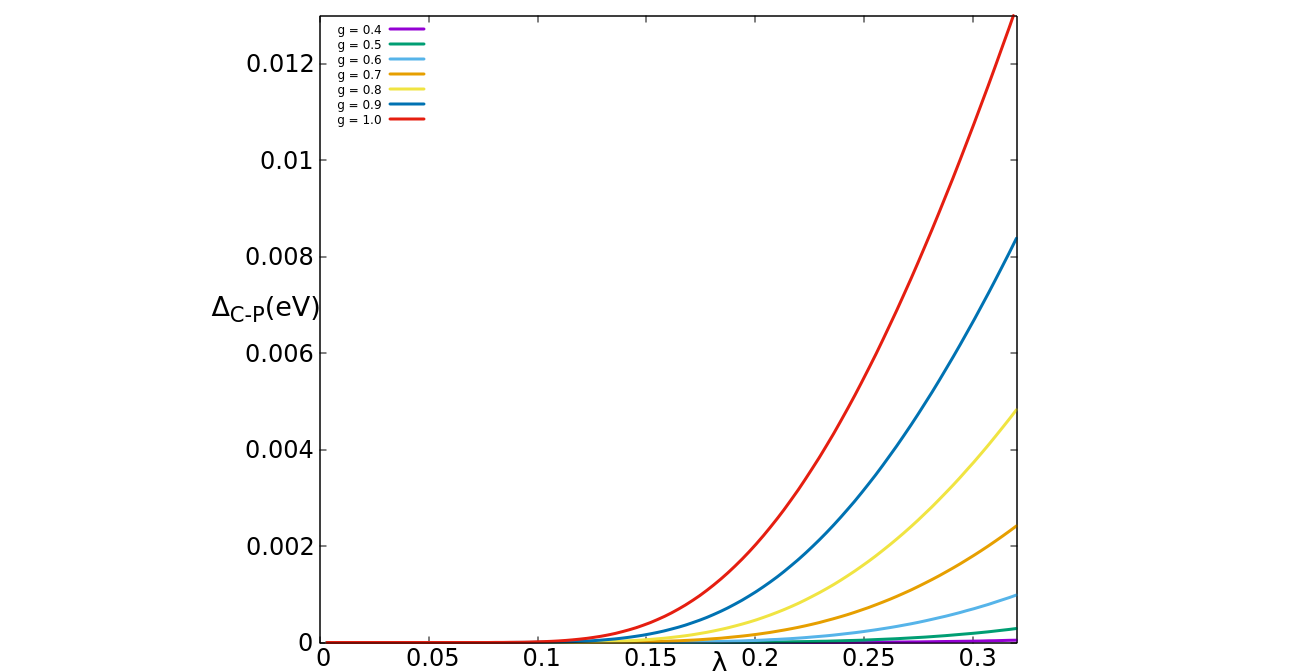}
\caption{$\Delta_{C-P}$ vs $\lambda $ for different g values}.\\
\end{figure}
In fig:4 above, where $\Delta_{C-P}$ is plotted against coupling constant for different Coulomb correlation strengths and it is observed that there exist a minimum value of g  to open up the pairing gap. The minimum cutoff values of g (denoted by $g_{min}$) corresponding to for the different $\lambda$ obtained from our calculations, are tabulated  below:

\begin{center}
	\begin{tabular}{ |p{6.0cm}|p{6.0cm}|  }
	\hline
	\multicolumn{2}{|c|}{Values of $g_{min}$  for different magnitudes of $\lambda$} \\
	\hline
\hspace*{2.8cm} $\lambda$ &  \hspace*{2.9cm} $g_{min}$ \\
	\hline
	\hspace*{2.8cm}0.15 & \hspace*{2.9cm}0.665\\
	\hspace*{2.8cm}0.20 & \hspace*{2.9cm}0.454\\
	\hspace*{2.8cm}0.25 & \hspace*{2.9cm}0.357\\
	\hspace*{2.8cm}0.30 & \hspace*{2.9cm}0.331\\
	\hline
	\end{tabular} \\
\end{center}
These values of $g_{min}$, give an indication of the available states for pairing in the presence of Coulomb correlation in system of our interest. As expected,  $g_{min}$ goes up as $\lambda$ decreases.
\section*{Conclusions:}
In this paper, we have presented the results of our investigation of the effect of the Coulomb correlation on the superconducting pairing of fermions, where the pairing mediator is a boson  and the repulsive correlation is introduced through Gutzwiller projection operator acting on an ideal BCS ground state. Our aim was to determine the upper bound of Coulomb correlation strength for which the pairing correlation can survive. We had adopted two distinct approaches involving both the (i) passive and (ii) `active' roles of Coulomb correlation parameter. The main highlights of our calculational results are given below:\\\\
(a) The physical characteristics of $\Delta_{C-P}$ is quite similar to that of $\Delta_{P} $ for the same range of the attractive potential well. \\
(b) The inter playing properties of pairing  correlation and Coulomb correlation in these two type superconducting gaps for a specified range of attractive potential, were presented earlier graphically in figs[1-4].  We now directly compare the dependence of these two types of gaps on `g' for a fixed $\lambda$ in the same plot [see fig:5]\\
\begin{figure}[H]
\centering
\includegraphics[scale=0.38]{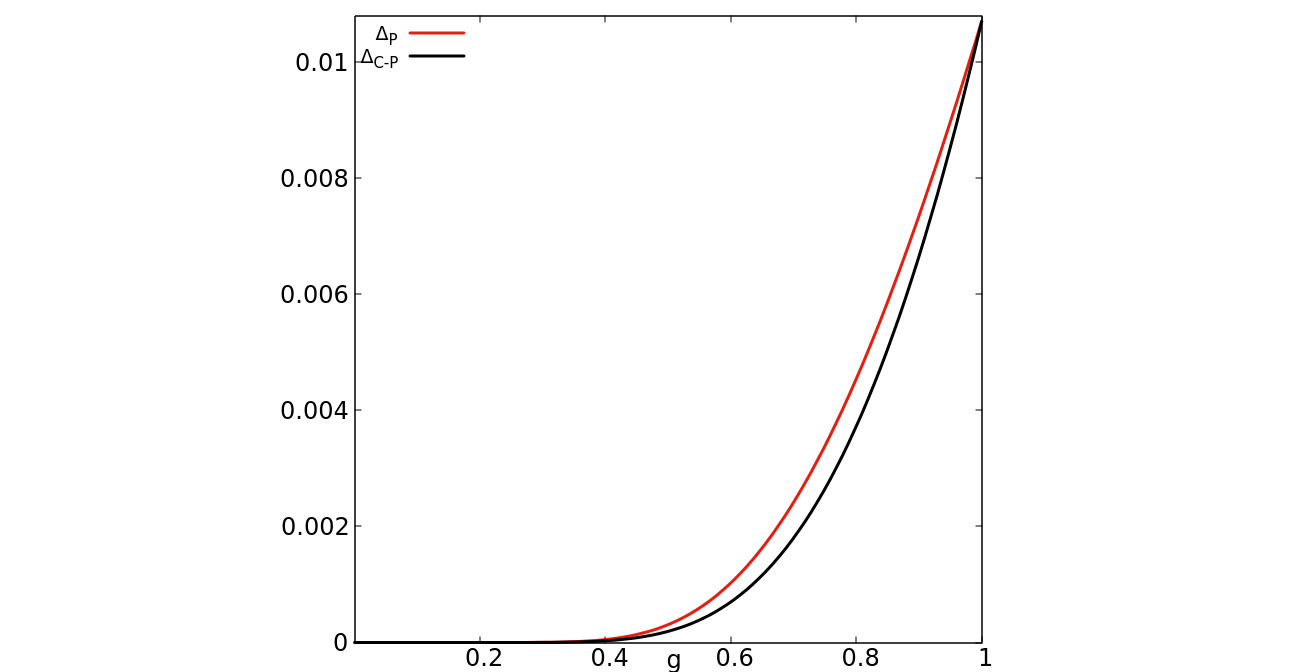}
\caption{$\Delta_{P}$ and $\Delta_{C-P}$ vs g  for coupling value $\lambda$ = 0.30 }
\end{figure}
The red and black lines are  for $\Delta_{P} $ and $\Delta_{C-P}$ respectively. Here, both the pairing gaps are calculated for a particular coupling strength of $\lambda$ (0.30). The inter playing nature of the two correlations is most prominent in the intermediate regime of g, where as both the pairing gaps exactly match at the two extremities besides their general overlap regions. Also, at the zero correlation i.e. g = 1,  both the pairing gaps ($\Delta_{P}$ and $\Delta_{C-P}$) are identical to the BSC pairing gap.  It is interesting to note that $ \Delta_{P} \geq \Delta_{C-P}$ in the entire regime of g values for a fixed $\lambda$.\\
(c) The pairing phenomena is studied in the weak coupling regime and we exclude the intermediate coupling(0.3 $\leqslant \lambda\leqslant $ 1) and strong coupling ($\lambda \geqslant $ 1) regime. In the strong coupling regime, there is a possibility of the formation of a composite electron-phonon quasi-particle and the calculation of $\lambda$ itself is quite non-trivial\cite{2,4,31}.\\
(d) The approach denoted by us as a case-I in the section: `Model system', where the superconducting pairing  itself originates from a Coulomb correlated Fermi sea, would be taken up in details soon and reported in a future publication.\\ 
\section*{Acknowledgment}
KM and RC would like to thank Department of Condensed Matter Physics and Material Sciences, S. N. Bose National Centre for Basic Sciences, for their kind hospitality. KM thanks UGC for the financial support.
\section*{Appendix-A}
\textbf{Normalization of the Coulomb correlated state:}
\vspace*{0.3 cm}
\begin{multline}
_{C}\langle\Psi\vert\Psi\rangle_{C} = \langle 0\vert\prod_{m} (u_{m}^{*} + v_{m}^{*}c_{-m,\downarrow}c_{m,\uparrow})\prod_{s}(1-\alpha \sum_{k^{'},m^{'}}c_{m^{'},\downarrow}^{+}c_{m^{'},\downarrow}c_{-k^{'},\uparrow}^{+}c_{-k^{'},\uparrow} e^{i(m^{'}-k^{'}).r_{s}})\\ \prod_{s^{'}}(1-\alpha \sum_{k^{''},m^{''}}c_{k^{''},\uparrow}^{+}c_{k^{''},\uparrow}c_{-m^{''},\downarrow}^{+}c_{m^{^{''},\downarrow}} e^{i(k^{''}-m^{''}).r_{s^{'}}})\prod_{l} (u_{l} + v_{l}c_{l,\uparrow}^{+}c_{-l,\downarrow}^{+})\vert 0\rangle \\
= \langle 0\vert \prod_{s,m}\prod_{s^{'},l}[u_{m}^{*}u_{l} + u_{m}^{*}v_{l} c_{l,\uparrow}^{+}c_{-l,\downarrow}^{+} - \alpha ( u_{m}^{*}u_{l}\sum_{k^{''},m^{''}} c_{k^{''},\uparrow}^{+} c_{k^{''},\uparrow} c_{-m^{''},\downarrow}^{+} c_{-m^{''},\downarrow} \\ u_{m}^{*}v_{l} \sum_{k^{''},m^{''}} c_{k^{''},\uparrow}^{+} c_{k^{''},\uparrow} c_{-m^{''},\downarrow}^{+} c_{-m^{''},\downarrow} c_{l,\uparrow}^{+}c_{-l,\downarrow}^{+})e^{i(k^{''}-m^{''}).r_{s^{'}}} + v_{m}^{*} u_{l} c_{-m,\downarrow}c_{m,\uparrow} \\ + v_{m}^{*} v_{l} c_{-m,\downarrow}c_{m,\uparrow} c_{l,\uparrow}^{+}c_{-l,\downarrow}^{+} - \alpha(
v_{m}^{*}u_{l}\sum_{k^{''},m{''}}c_{m,\downarrow}c_{m,\uparrow}c_{k^{''},\uparrow}^{+}c_{k^{''},\uparrow} c_{-m^{''},\downarrow}^{+} c_{-m^{''},\downarrow}  \\ +
v_{m}^{*} v_{l} \sum_{k^{''},m{''}}c_{m,\downarrow}c_{m,\uparrow}c_{k^{''},\uparrow}^{+}c_{k^{''},\uparrow} c_{-m^{''},\downarrow}^{+} c_{-m^{''},\downarrow} c_{l,\uparrow}^{+} c_{-l,\downarrow}) e^{i(k^{''}-m^{''}).r_{s^{'}}} \\ -\alpha (u_{m}^{*}u_{l} \sum_{k^{'},m^{'}} c_{m^{'},\downarrow}^{+}c_{m^{'},\uparrow} c_{-k^{'},\uparrow}c_{-k^{'},\uparrow} + u_{m}^{*}v_{l}\sum_{k^{'},m^{'}} c_{m^{'},\downarrow}^{+}c_{m^{'},\uparrow} c_{-k^{'},\uparrow}c_{-k^{'},\uparrow} c_{l,\uparrow}^{+}c_{-l'\downarrow}^{+})e^{-i(k^{'}-m^{'}).r_{s}} \\
+ \alpha^{2} (u_{m}^{*}u_{l}\sum_{k^{'},m^{'},k^{''},m{''}} c_{m^{'},\downarrow}^{+}c_{m^{'},\downarrow} c_{-k^{'},\uparrow}^{+}c_{-k^{'},\uparrow} c_{k^{''},\uparrow}^{+}c_{k^{''},\uparrow} c_{-m^{''},\downarrow}^{+}c_{-m^{''},\downarrow} e^{i(k^{''}-m^{''}).r_{s^{'}} -(k^{'}-m^{'}).r_{s}} + \\ u_{m}^{*}v_{l}\sum_{k^{'},m^{'},k^{''},m{''}} c_{m^{'},\downarrow}^{+}c_{m^{'},\downarrow} c_{-k^{'},\uparrow}^{+}c_{-k^{'},\uparrow} c_{k^{''},\uparrow}^{+}c_{k^{''},\uparrow} c_{-m^{''},\downarrow}^{+}c_{-m^{''},\downarrow} c_{l,\uparrow}^{+} c_{-l,\downarrow}^{+} e^{i(k^{''}-m^{''}).r_{s^{'}} -(k^{'}-m^{'}).r_{s}}) \\ - \alpha( v_{m}^{*}u_{l}\sum_{k^{'},m^{'}} c_{-m,\downarrow} c_{m,\uparrow} c_{m^{'},\downarrow}^{+}c_{m^{'},\downarrow}c_{-k^{'},\uparrow}^{+}c_{-k^{'},\uparrow} e^{i(k^{'}-m^{'}).r_{s}} + \\ v_{m}^{*}v_{l} \sum_{k^{'},m^{'}}c_{-m,\downarrow}c_{-m,\uparrow}c_{m^{'},\downarrow}^{+}c_{m^{'},\downarrow}c_{-k^{''},\uparrow}^{+}c_{-k^{''},\uparrow}c_{l,\uparrow}^{+}c_{-l,\downarrow}^{+} e^{-i(k^{'}-m^{'}).r_{s}}) + \\
 \alpha^{2}( v_{m}^{*}u_{l} \sum_{k^{'},m^{'},k^{''},m^{''}} c_{-m,\downarrow}c_{-m,\uparrow}c_{m^{'},\downarrow}^{+}c_{m^{'},\downarrow}c_{-k^{'},\uparrow}^{+}c_{-k^{'},\uparrow} c_{k^{''},\uparrow}^{+}c_{k^{''},\uparrow} c_{-m^{''},\downarrow} c_{-m^{''},\downarrow} \\ e^{i((k^{''}-m^{''})r_{s^{'}}-(k^{'}-m^{'})r_{s})} \\ +
 v_{m}^{*}v_{l} \sum_{k^{'},m^{'},k^{''},m^{''}} c_{-m,\downarrow}c_{-m,\uparrow}c_{m^{'},\downarrow}^{+}c_{m^{'},\downarrow}c_{-k^{'},\uparrow}^{+}c_{-k^{'},\uparrow} c_{k^{''},\uparrow}^{+}c_{k^{''},\uparrow} c_{-m^{''},\downarrow}^{+} c_{-m^{''},\downarrow} c_{l,\uparrow}^{+} c_{-l,\downarrow}^{+} \\e^{i((k^{''}-m^{''})r_{s^{'}}-(k^{'}-m^{'})r_{s})})]\vert 0 \rangle\\
\end{multline}
Here, the orthogonality of the states set a  condition, for which only the terms with equal number of fermion Creation and annihilation operators will contribute with  a non-zero value to that normalization. Therefore,
\begin{multline}
_{C}\langle\Psi\vert\Psi\rangle_{C} = \langle 0\vert \prod_{s,m}\prod_{s^{'},l}[ u_{m}^{*}u_{l} + v_{m}^{*}v_{l} c_{-m,\downarrow}c_{m,\uparrow} c_{l,\uparrow}^{+}c_{-l,\downarrow}^{+}\\ -\alpha v_{m}^{*}v_{l} \sum_{k^{''},m^{''}} c_{-m,\downarrow}c_{m,\uparrow} c_{k^{''},\uparrow}^{+}c_{k^{''},\uparrow}c_{-m^{''},\downarrow}^{+}c_{-m^{''},\downarrow} c_{l,\uparrow}^{+}c_{-l,\downarrow}^{+}
e^{i(k^{''}-m^{''}).r_{s^{'}}} \\
+ \alpha^{2} v_{m}^{*}v_{l} \sum_{k^{'},m^{'},k^{''},m^{''}} c_{-m,\downarrow}c_{m,\uparrow} c_{m^{'},\downarrow}^{+}c_{m^{'},\downarrow}c_{-k^{'},\uparrow}^{+}c_{-k^{'},\uparrow} c_{k^{''},\uparrow}^{+}c_{k^{''},\uparrow}c_{-m^{''},\downarrow}^{+}c_{-m^{''},\downarrow} c_{l,\uparrow}^{+}c_{-l,\downarrow}^{+}\\
e^{i((k^{''}-m^{''}).r_{s^{'}}-(k^{'}-m^{'}).r_{s})} \\
-\alpha v_{m}^{*}v_{l} \sum_{k^{'},m^{'}}c_{-m,\downarrow}c_{m,\uparrow} c_{-m^{'},\downarrow}^{+} c_{-m^{'},\downarrow} c_{-k^{'},\uparrow}^{+}c_{-k^{'},\uparrow} c_{l,\uparrow}^{+}c_{-l,\downarrow}^{+} e^{i(k^{'}-m^{'}).r_{s}}] \vert 0\rangle \\
 = \prod_{s,m}\prod_{s^{'},l} [ u_{m}^{*}u_{l} + v_{m}^{*}v_{l} - \alpha v_m^{*}v_{l}\delta_{-l,-m^{''}} \delta_{l,k^{''}} \delta_{-m^{''},-m} \delta_{k^{''},m}    + \\
 \alpha^{2} v_{m}^{*}v_{l}\delta_{k^{''},m^{''}} \delta_{k^{'},m^{'}} - \alpha v_m^{*}v_{l} \delta_{-m^{'},-l}\delta_{l,k^{'}} \delta_{-k^{'},m}\delta_{m^{'},m}]\\
\end{multline}
\begin{align*}
 _{C}\langle\Psi\vert\Psi\rangle_{C} = \prod_{l} [ \vert u_{l}\vert^{2}  + \vert v_{l}\vert^{2}-2\alpha \vert v_{l} \vert^{2} + \alpha^{2} \vert v_{l}\vert^{2}] \\
 _{C}\langle\Psi\vert\Psi\rangle_{C} = \prod_{l} [1+(\alpha^{2}-2\alpha)\vert v_{l}\vert^{2}] ;[\because \vert u_{k}\vert^{2} + \vert v_{k}\vert^{2} = 1 ]
\end{align*}
\section*{Appendix-B}
\textbf{Total energy expectation W:}
\begin{equation}
W   = \dfrac{1}{_{c}\langle\Psi\vert\Psi\rangle_{c}} \left[  { _{c}\langle\Psi\vert \sum_{k} 2\varepsilon_{k} b^{+}_{k}b_{k}  +  \sum_{k,l} V_{k,l}b^{+}_{k} b_{l}
\vert\Psi\rangle_{c} }\right]
\end{equation}
Kinetic energy operator expectation value:
\begin{eqnarray*}
T = \dfrac{1}{\prod_{l} [1+(\alpha^{2}-2\alpha)\vert v_{l}\vert^{2}]} \sum_{k,\sigma} \varepsilon_{k} \langle 0\vert\prod_{m} (u_{m}^{*} + v_{m}^{*}c_{-m,\downarrow}c_{m,\uparrow})\prod_{s}(1-\alpha \sum_{k^{'},m^{'}}c_{m^{'},\downarrow}^{+}c_{m^{'},\downarrow}c_{-k^{'},\uparrow}^{+}c_{-k^{'},\uparrow}\\ e^{i(m^{'}-k^{'}).r_{s}}) c_{k,\sigma}^{+}c_{k,\sigma}\prod_{s^{'}}(1-\alpha \sum_{k^{''},m^{''}}c_{k^{''},\uparrow}^{+}c_{k^{''},\uparrow}c_{-m^{''},\downarrow}^{+}c_{m^{^{''},\downarrow}} e^{i(k^{''}-m^{''}).r_{s^{'}}})\prod_{l} (u_{l} + v_{l}c_{l,\uparrow}^{+}c_{-l,\downarrow}^{+})\vert 0\rangle
\end{eqnarray*}
This calculation is done by classifying the contribution coming out from the states which are specified with the power of $\alpha$, and they are listed as below:
\begin{equation}
\alpha^{0} : 2 \sum_{k} \dfrac{\varepsilon_{k} \vert v_{k}\vert^{2}}{1+\alpha(\alpha-2)\vert v_{k}\vert^{2}} 
\end{equation}
\begin{equation}
\alpha^{1} : 2 \sum_{k} \dfrac{-2\alpha\varepsilon_{k} \vert v_{k}\vert^{2}}{1+\alpha(\alpha-2)\vert v_{k}\vert^{2}} 
\end{equation}
\begin{equation}
\alpha^{2} : 2 \sum_{k} \dfrac{\alpha^{2}\varepsilon_{k} \vert v_{k}\vert^{2}}{1+\alpha(\alpha-2)\vert v_{k}\vert^{2}} 
\end{equation}
The contribution of potential energy expectation value is deduced in a similar manner and these are added in this section.
\begin{equation}
\alpha^{0} : 0
\end{equation} 
\begin{equation}
\alpha^{1} : \sum_{k,l} V_{k,l}\dfrac{-2\alpha u_{l}^{*}u_{k}v_{k}^{*}v_{l}}{[1+\alpha(\alpha-2)\vert v_{k}\vert^{2}][1+\alpha(\alpha-2)\vert v_{l}\vert^{2}]} 
\end{equation}  
\begin{equation}
\alpha^{2} : \sum_{k,l} V_{k,l}\dfrac{(1+\alpha^{2}) u_{l}^{*}u_{k}v_{k}^{*}v_{l}}{[1+\alpha(\alpha-2)\vert v_{k}\vert^{2}][1+\alpha(\alpha-2)\vert v_{l}\vert^{2}]} 
\end{equation} 
\bibliographystyle{unsrt}

\end{document}